\documentclass[12pt,showkeys,nofootinbib,onecolumn,prX,letter,secnumarabic,amsmath,tightenlines]{revtex4}
\usepackage{amsmath,amssymb,amsthm}
\usepackage{bbm}
\usepackage{dsfont}
\usepackage[colorlinks=true, pdfstartview=FitH, linkcolor=blue,citecolor=blue, urlcolor=blue,pdftex]{hyperref}
\usepackage[latin1]{inputenc}

\newcommand{\Cx}{{\mathbb C}}

\newcommand{\idty}{\mathbbm{1}}
\DeclareMathOperator{\id}{id}

\DeclareMathOperator{\tr}{Tr}
\newcommand{\<}{\langle}
\renewcommand{\>}{\rangle}

\providecommand{\ket}[1]{|#1\rangle}
\providecommand{\bra}[1]{\langle#1|}

\renewcommand{\c}[1]{\mathcal{#1}}
\newcommand{\g}[1]{\mathfrak{#1}}
\newcommand{\s}[1]{\mathsf{#1}}
\renewcommand{\r}[1]{\mathrm{#1}}

\renewcommand{\det}{\r{Det}\,}
\DeclareMathOperator*{\loplus}{\mbox{\Large\mbox{$\oplus$}}}
\newtheorem{lemma}{Lemma}
\newtheorem{proposition}{Proposition}

\theoremstyle{remark}
\newtheorem{remark}{Remark}
\newcommand{\A}{\mathfrak{A}}

\hypersetup{
    pdftitle={Fermionic Quasi-free States and Maps in Information Theory},
    pdfauthor={B.~Dierckx, M.~Fannes and M.~Pogorzelska},
    pdfkeywords={quasi-free states, fermions, completely positive maps, Choi matrix, Jamiolkowski state, entropy}
}

\begin{document}

\title{Fermionic Quasi-free States and Maps in Information Theory}
\author{B.~Dierckx}
\email{brecht.dierckx@fys.kuleuven.be}
\author{M.~Fannes}
\email{mark.fannes@fys.kuleuven.be}
\affiliation{Instituut voor Theoretische Fysica, KULeuven}
\author{M.~Pogorzelska}
\email{dokmpo@univ.gda.pl}
\affiliation{Institute of Theoretical Physics and Astrophysics, University of Gda\'{n}sk}

\begin{abstract}

This paper and the results therein are geared towards building a basic toolbox for calculations in quantum information theory of quasi-free fermionic systems.  Various entropy and relative entropy measures are discussed and the calculation of these reduced to evaluating functions on the one-particle component of quasi-free states.

The set of quasi-free affine maps on the state space is determined and fully characterized in terms of operations on one-particle subspaces.  For a subclass of trace preserving completely positive maps and for their duals, Choi matrices and Jamiolkowski states are discussed.

\end{abstract} 

\maketitle

\section{Introduction}

There are not too many classes of states or quantum operations that can be handled in detail. Well-known examples are gaussian structures in bosonic systems \cite{BratRob2} and exchangeable states and maps for spin systems \cite{Caves2002,Hudson1976}. Although the systems under consideration can often be quite large, even infinite, the computational difficulty of most associated quantities is many orders of magnitude lower than what one encounters in more general systems.  In particular gaussian states have been used in quantum optics to that effect  for many years.  Perhaps because of the close link between the two fields, gaussian states were also the first of the aforementioned classes to come under consideration in quantum information theory.  Recent years have seen a large amount of work done on their role in bosonic systems, see for instance~\cite{Serafini2005}.

This paper deals with quasi-free fermionic systems \cite{BratRob2}. In field theory and statistical mechanics such effective free evolutions and states have been used extensively as an approximation to interacting systems, a well-known example being the Hartree-Fock approximation. The main simplifying feature lies in the particular combinatorial properties of correlation functions and maps. In fact, states and maps are fully determined by one-particle operators. As the dimension of the observables increases exponentially with the dimension of the one-particle space we obtain a very significant reduction of complexity.

Fermionic systems and quasi-free states should be of particular interest to information theorists.  A qubit system can always be mapped on an interacting system of fermions by the so-called \textbf{Jordan-Wigner isomorphism}.  A particular subset of quantum operations on qubits can then be identified with quasi-free evolutions of fermionic systems and in these problems quasi-free states play an important role. A recent article dealing with this duality between fermions and qubits is \cite{Ovrum2007}.

We attempted to write a rather self-contained paper, offering a toolbox for computations and testing ideas in quantum information. Exponential elements instead of the standard creation and annihilation operators, see~\cite{Derezinski2006}, appear to be an efficient computational tool. Well-known objects, as quasi-free states, are reconsidered in these terms but the main goal are quasi-free quantum operations. As we restrict ourselves on purpose to finite dimensions we only need linear algebra. This limitation can be overcome: infinite fermionic systems no longer have a canonical representation and so one has to use appropriate representations. This typically involves introducing trace-class conditions on the one-particle operators.

The paper is organized as follows: \autoref{sec:fock} recalls some basic properties of fermionic Fock space and creation and annihilation operators satisfying CAR (canonical anticommutation relations). In \autoref{sec:GICAR} we introduce the basic exponential operators and study their properties. \hyperref[sec:states]{Section~\ref*{sec:states}} reconsiders quasi-free states and introduces new calculation techniques for some entropic quantities. \hyperref[sec:maps]{Section~\ref*{sec:maps}} deals with quasi-free quantum operations.

\section{Fermionic Fock Space}
\label{sec:fock}

The quantum mechanical description of fermions is firmly connected to the mathematical concept of  antisymmetric Fock space, especially in finite dimensions.  Throughout the article we will regularly fall back on this to prove, calculate or physically motivate our expressions.

In the following, $\g H$ will denote a finite dimensional complex inner product space, which to almost all intent and purpose,  we can regard as a finite Hilbert space.  Many of the results can be extended to infinite dimensional Hilbert spaces, modulo, of course, some suitable additional conditions.

The symbol $\otimes$ is well-known as a notation for the tensor product of two or more objects.  In this, we will use the symbol $\wedge$ to denote the \emph{antisymmetric} tensor product of vectors, operators and even algebras. Although the meaning of the symbol will change depending on the setting in which it is used, restricting ourselves to a single symbol, greatly simplifies the notation and looks, at least to a physicist's eye, more elegant.  As is often the case with degenerate notations, the context should specify which version of the wedge we are talking about.

\subsection{Antisymmetric Vector Spaces}

We define the $k$-fold antisymmetric tensor product or wedge product of vectors $\varphi_1$,~$\varphi_2$,~\ldots,~$\varphi_k$ in $\g H$ as
\begin{equation*}
\varphi_1 \wedge \varphi_2 \wedge \ldots \wedge \varphi_k := \frac{1}{\sqrt{k!}} \sum_\sigma \epsilon(\sigma) \, \varphi_{\sigma(1)} \otimes \varphi_{\sigma(2)} \otimes \ldots \otimes \varphi_{\sigma(k)}
\end{equation*}
where $\sigma$ runs over all permutations of the $k$ indices and $\epsilon(\sigma) = \pm$ depending on the parity of the permutation, $+$ if even, $-$ if odd.

Let $U_{\sigma}$ be the unitary operator which implements the permutation $\sigma$ on $\otimes^k \g H$. The fully antisymmetric subspace of $\otimes^k \g H$ consists of these vectors $\eta \in \otimes^k \g H$ which satisfy $U_\sigma \, \eta = \epsilon(\sigma) \eta$. It is spanned by the $k$-fold antisymmetric vectors and we denote it as $\wedge^k \g H$ or $\g H^{(k)}$. This space has the same inner product as $\otimes^k \g H$, which can be written more succinctly as
\begin{equation*}
\bigl\< \varphi_1 \wedge \varphi_2 \wedge \cdots \wedge \varphi_k \, ,\, \psi_1 \wedge \psi_2 \wedge \cdots \wedge \psi_k \bigr\>
= \det\Bigl( \bigl[ \<\varphi_i \, ,\, \psi_j\> \bigr]_{i,j} \Bigr).
\end{equation*}
This in turn makes it quite easy to transport a basis $\bigl\{ e_1, e_2, \ldots \bigr\}$ of $\g H$ to $\g H^{(k)}$. To each such orthonormal basis corresponds an orthonormal basis
\begin{equation} \label{asymbasis}
\bigl\{ e_\Lambda \, :\, \Lambda \subset \{1,2,\ldots\}\ \&\ \#(\Lambda) = k \bigr\}
\end{equation}
of $\g H^{(k)}$ where
\begin{equation*}
e_\Lambda := e_{i_1} \wedge e_{i_2} \wedge \cdots \wedge e_{i_k}, \qquad i_1 < i_2 < \cdots < i_k\ \&\ \Lambda = \{i_1, i_2, \ldots, i_k \}.
\end{equation*}

The wedge product extends naturally to antisymmetric vector spaces. If $\varphi \in \g H^{(k)}$ and $\chi \in \g H^{(\ell)}$ then
\begin{equation*}
\varphi \wedge \chi := \frac{1}{\sqrt{\binom{k+\ell}{k}}}\  \sum \epsilon(\Lambda,M) \varphi_\Lambda \otimes \chi_M
\end{equation*}
is an element of  $\g H^{(h+\ell)}$. The sum runs over the ordered partitions of $\{\, 1,2,\ldots, k+\ell\, \}$ in subsets $\Lambda$ and $M$ with $\#(\Lambda) = k$ and $\#(M) = \ell$, $\varphi_\Lambda$ and $\chi_M$ are the ordered injections of $\varphi$ and $\chi$ in the corresponding tensor products and $\epsilon(\Lambda,M)$ is the parity of the permutation $(\Lambda, M)$. To clarify this a bit, assume that $\varphi$ and $\chi$ are elementary antisymmetric tensors, i.e.\
\begin{equation*}
\varphi = \varphi_1 \wedge \ldots \wedge \varphi_k \qquad\text{and}\qquad \chi = \chi_1 \wedge \ldots \wedge \chi_{\ell}.
\end{equation*}
Then there is a canonical way to define the wedge product between these two vectors as
\begin{equation*}
\begin{split}
\varphi \wedge \chi
&= \varphi_1 \wedge \ldots \wedge \varphi_k \wedge \chi_1 \wedge \ldots \wedge \chi_{\ell}\\
&= \frac{1}{\sqrt{(k+\ell)!}} \sum_{\sigma} \epsilon(\sigma) U_{\sigma} \bigl( \varphi_1 \otimes \ldots \otimes \varphi_k \otimes \chi_1  \otimes \ldots \chi_{\ell} \bigr).
\end{split}
\end{equation*}
The above defined extension to wedge products of general antisymmetric vectors is then just the linear extension of this and so it becomes easy to prove the following important property of the wedge.

\begin{lemma}
The wedge operation is associative
\begin{equation*}
 (\varphi \wedge \chi) \wedge \psi
 = \varphi \wedge (\chi \wedge \psi).
\end{equation*}
\end{lemma}

\subsection{Construction of the Fermionic Fock Space}

Although the $k$-antisymmetric vectors we introduced in the previous section are an apt description for the states or wave functions of $k$ fermionic particles, we need more to properly describe a physical system of such particles. In nature there are processes which do not conserve the number of particles.  So we need a setting in which we can jump between different $k$-antisymmetric vector spaces.

Consider for example a fermion with $d$ modes. The state space associated with it is then also $d$-dimensional. If we add another identical particle to our system, the degrees of freedom do not go up as they do with qubits, rather, since we need to obey the Pauli exclusion principle, the dimension of the state space, which is the $2$-antisymmetric space, is only $\binom{d}{2}$. In general, the dimension of the state space of $k$ such particles is $\binom{d}{k}$ as can be readily seen from~(\ref{asymbasis}), and in particular, the state space of $d$ such particles is only one-dimensional. So we can only combine $d$ fermions before we literally run out of space to put them.

To unify these concepts of particle creation/annihilation and the exclusion principle, the antisymmetric Fock space is introduced. The fermionic Fock space with one-particle space $\g H$ is
\begin{equation*}
\Gamma(\g H) := \Cx \oplus \g H \oplus \g H^{(2)} \oplus \cdots \oplus \Cx.
\end{equation*}
It follows from the paragraph above that $\Gamma(\c H)$ has dimension $2^d$ where $d = \dim(\g H)$. The first term in the direct sum is the vacuum state and the last one is the completely filled Fermi sea.

\subsubsection{Piecing two systems together}

Given an orthogonal decomposition of $\g H$ into subspaces $\g H_1$ and $\g H_2$,
there is a natural isomorphism between $\Gamma(\g H)$ and the tensor
product of the Fock spaces with $\g H_1$ and $\g H_2$ as one-particle
spaces
\begin{equation*}
\Gamma\bigl( \g H_1 \oplus \g H_2 \bigr)
\cong \Gamma\bigl( \g H_1 \bigr) \otimes \Gamma\bigl( \g H_2 \bigr)
\end{equation*}
explicitly given by
\begin{equation*}
(\varphi_1 \oplus \psi_1) \wedge \cdots \wedge (\varphi_k \oplus \psi_k) \
\cong {\textstyle\sum^\oplus} \bigl( \varphi_{i_1} \wedge \cdots \wedge \varphi_{i_r} \bigr) \otimes \bigl( \psi_{j_1} \wedge \cdots \wedge \psi_{j_s} \bigr).
\end{equation*}
The summation sign $\sum^\oplus$ points at mixed sums and direct sums and the summation runs over all ordered partitions of $\{1,2,\ldots,k\}$ in two subsets $\{ i_1, i_2, \ldots, i_r\}$ and $\{j_1, j_2, \ldots, j_s\}$. Due to the antisymmetry one has to pay attention to the order of the factors in this isomorphism.

\subsubsection{Elementary antisymmetric vectors}

We would also like to point out a peculiar property of elementary vectors which will come in handy later on.

\begin{lemma}
A nonzero vector $\varphi \in \g H^{(k)}$ is an elementary vector, i.e.\ can be written as
\begin{equation*}
\varphi = \psi_1 \wedge \ldots \wedge \psi_k
\end{equation*}
if and only if the space
\begin{equation*}
\{ \chi \in \g H \, | \,  \chi \wedge \varphi = 0 \}
\end{equation*}
is $k$-dimensional.
\end{lemma}

\begin{proof}
Consider an elementary vector $\varphi$ in $\g H^{(k)}$ which can be written in the simple form
\begin{equation*}
\varphi = \psi_1 \wedge \ldots \wedge \psi_k
\end{equation*}
where $\{ \psi_i\}_i \cup \{ \chi_\ell\}_\ell$ forms a basis for the generating space $\g H$ and $\{ \psi_i\}_i \bot \{ \chi_\ell\}_\ell$.  From the above, it is clear that a set of vectors is linearly dependent if and only if the elementary tensor constructed from them is zero.  The $\chi_\ell$ are linearly independent of the $\psi_i$ and so cannot contribute to the set in the lemma. Only vectors which are built up exclusively out of the $\psi_i$ contribute and the space generated by the $\psi_i$ is $k$-dimensional.

Now assume that the set mentioned in the lemma is indeed $k$-dimensional. This implies that we can find $(d-k)$ orthogonal vectors in the $d$-dimensional $\g H$ which are linearly independent of this set and thus of the constituting vectors of $\varphi$. So only $k$ linearly independent vectors can be involved in the creation of $\varphi$. But the only non-zero $k$-antisymmetric vectors which can be built out of $k$ vectors, are proportional to each other and elementary.
\end{proof}

\subsection{The CAR Algebra}

Fermionic Fock space can also be built in another way. Any vector $\varphi$ of $\g H$ induces a linear operator $a^*(\varphi)$ from $\g H^{(k)}$ to $\g H^{(k+1)}$.  We first define the action of $a^*(\varphi)$ on elementary vectors and then linearly extend it to the whole space.
\begin{equation*}
a^*(\varphi) (\psi_1 \wedge \ldots \wedge \psi_k)
:= \varphi \wedge \psi_1 \wedge \ldots \wedge \psi_k.
\end{equation*}
The operator $a^*(\varphi)$ is called a creation operator, its adjoint an annihilation operator and in effect they emulate the creation or destruction of a Fermion in the state $\varphi$.  By repeatedly applying creation operators to the vacuum vector, we can build up the entire Fock space.

Although we will not use this language very frequently, the operators defined above satisfy exactly the CAR required for a quantum mechanical description of Fermions
\begin{equation} \label{CAR}
\{a(\varphi),a(\psi)\} = 0 \qquad \text{and}\qquad \{a(\varphi),a^*(\psi)\} = \< \phi \,,\, \psi \> \, \idty.
\end{equation}

The algebra built on the creation and annihilation operators is called the CAR-algebra in reference to the important commutation relations~(\ref{CAR}). It coincides with the algebra of linear transformations of the Fock space $\Gamma(\g H)$. This algebra is in fact a universal algebra because it is the unique algebra generated by a unit element $\idty$ and by $\{ a(\varphi) \,:\, \varphi \in \g H \}$ such that the operators $a(\varphi)$ satisfy the CAR-conditions and that the map $\varphi \mapsto a(\varphi)$ is complex antilinear. We will denote it by $\g A(\g H)$.

The following proposition will be needed later on. It is well-known, so we state it without proof, see~\cite{BratRob2}.

\begin{proposition} \label{prop:wedgestate}
Suppose that $\omega$ is an even state, i.e.\ vanishes on monomials in creation and annihilation operators with odd number of factors, on $\g A(\g H)$ and that $\sigma$ is a state on $\g A(\g K)$; then there exists a unique state $\omega \wedge \sigma$ on $\g A(\g H) \wedge \g A(\g K):=\g A(\g H \oplus \g K) $, defined by
\begin{equation*}
(\omega \wedge \sigma) \, (x y)
:= \omega(x) \, \sigma(y), \qquad x \in \g A(\g H), \, y \in \g A(\g K).
\end{equation*}
\end{proposition}

\begin{remark}
The wedge product $\g A(\g H) \wedge \g A(\g K)$ we implicitly defined in the above proposition is not the same as the tensor product of the two algebras as can be seen from the following constructive explanation.

There is a natural embedding
\begin{equation*}
  \jmath_1: \A(\g H_1)\hookrightarrow \A(\g H_1\oplus\g H_2):
  \jmath_1(a(\varphi_1))= a(\varphi_1\oplus 0), \quad \varphi_1\in\g H_1
\end{equation*}
and of course an analogous embedding $\jmath_2$ of $\A(\g H_2)$. Clearly, $\jmath_1(\A(\g H_1))$ and $\jmath_2(\A(\g H_2))$ generate $\A(\g H_1\oplus\g H_2)$ but they do not sit in $\A(\g H_1\oplus\g H_2)$ as tensor factors because
\begin{equation*}
  \{a^\#(\varphi_1\oplus 0) \, ,\, a^\#(0\oplus\varphi_2)\}= 0
  \quad\textrm{instead of}\quad
  [a^\#(\varphi_1\oplus 0) \, ,\, a^\#(0\oplus\varphi_2)]= 0,
\end{equation*}
$a^\#$ denotes either $a$ or $a^*$. Sometimes, $\A(\g H_1\oplus\g H_2)$ is called the graded tensor product of $\A(\g H_1)$ and $\A(\g H_2)$.
\end{remark}

\section{The GICAR Algebra}
\label{sec:GICAR}

For any one-particle basis $\{ e_i\}$, a special operator can be defined
\begin{equation} \label{numberoperator}
N:=\sum_i a^*(e_i) a(e_i).
\end{equation}
It is invariant under the gauge group of $\g A (\g H)$, but also under any arbitrary basis transformation of the one-particle space and so~(\ref{numberoperator}) defines a unique operator in the algebra.
Consider the action of $N$ on an arbitrary (non-zero) $k$-particle vector
\begin{equation*}
N\, \varphi_1 \wedge \ldots \wedge \varphi_k
= k \, \varphi_1 \wedge \ldots \wedge \varphi_k.
\end{equation*}
So, $N$ counts the number of particles in a given state and as such is called the number operator. Its eigenspaces are obviously the $k$-antisymmetric spaces and its spectrum consists of the integers $\{0, \ldots, d\}$. The commutant of the number operator, is called the gauge invariant CAR-algebra or GICAR for short. It is the largest subalgebra of $\g A(\g H)$ that is invariant under the gauge group. It is also generated as the span of all monomials in $a$,$a^*$ containing as many $a$'s as $a^*$'s.

\subsubsection{Exponential elements}

With $d = \dim(\g H)$ and $\c M_k$ the algebra of complex square matrices of dimension $k$, the GICAR-algebra can be written as
\begin{equation*}
\g A^{\r{GICAR}}(\g H) = \Cx \oplus \c M_d \oplus \c M_{d(d-1)/2} \oplus \cdots \oplus \Cx.
\end{equation*}
The dimension of $\g A^{\r{GICAR}}(\g H)$, counted as a complex vector space, is then seen to be $\binom{2d}{d}$.

Remark that the GICAR is the subalgebra of the CAR of block diagonal transformations of Fock space, in particular it contains elements of the form
\begin{equation*}
\s E(X) := 1 \oplus X \oplus (X \otimes X)\bigr|_{\g H(2)} \oplus \cdots \oplus (\otimes^d X ) \bigr|_{\g H(d)}
\end{equation*}
or with an obvious notational meaning
\begin{equation*}
\s E(X) = 1 \oplus X \oplus (X \wedge X) \oplus \cdots \oplus (\wedge^d X).
\end{equation*}
Their spectrum $\sigma \bigl(\s E(X)\bigr)$ can be computed to be
\begin{equation*}
\sigma\bigl( \s E(X) \bigr) = \Bigl\{ \prod_{i \in \Lambda} \lambda_i \,:\, \lambda_i \in \sigma(X) \, \& \, \Lambda \subset  \{1, \ldots, \text{Rank}(X)\} \Bigr\}
\end{equation*}
This property is easily verified by looking at the antisymmetric part of $\otimes_k X$.  As a tensor product, the eigenvalues of this are exactly all possible monomials of the eigenvalues of $X$ of length $k$. If we number the eigenvalues of $X$ repeated according to their multiplicities as $\lambda_i$, then a monomial of the $\lambda_i$ will contribute to the spectrum of $\wedge_k X$ if and only if each $\lambda_i$ appears exactly once or not at all.

They also enjoy the following properties
\begin{align}
&\s E(\idty) = \idty
\nonumber \\
&\s E(X)^* = \s E(X^*)
\nonumber \\
&\s E(X) \s E(Y) = \s E(XY)
\nonumber \\
&\s E(X) \ge 0 \text{ iff } X \ge 0
\nonumber \\
&\s E\bigl( X_1 \oplus X_2 \bigr) \cong \s E\bigl(X_1 \bigr) \otimes \s E\bigl(X_2 \bigr)
\nonumber \\
&\tr \s E(X) = \det(\idty + X)
\nonumber \\
&\g A^{\r{GICAR}}(\g H) = \r{Span}\bigl( \{\s E(X) \}\bigr).
\label{Eprops}
\end{align}
The first six properties can be easily checked by looking at the spectrum of the $\s E$-operators and the isomorphism we mentioned above. The last property follows from expanding $\lambda \mapsto \s E(\lambda X)$ around $\lambda = 0$ and remarking that
$\r{Span} \bigl( \bigl\{ \otimes^k X \bigr|_{\g H^{(k)}} \bigr\} \bigr)$ coincides with the linear transformations of $\g H^{(k)}$.

\subsubsection{$k$-Particle projectors}

Another useful set of elements consists of certain one-dimensional projectors. Given a $k$-dimensional subspace $\g K$ of $\g H$, all vectors $\varphi_1 \wedge \cdots \wedge \varphi_k$ with $\varphi_1,\ldots, \varphi_k \in \g K$ are proportional and span therefore a one-dimensional subspace of $\g H^{(k)}$. The projector on that space will be denoted by $\s P_\ast(\g K)$. If we denote by $[\g K]$ the projector on $\g K$ then
\begin{equation*}
\s P_\ast(\g K)
= [\g K] \otimes \cdots \otimes [\g K] \Bigr|_{\g H^{(k)}}.
\end{equation*}
It is convenient to associate the projector on the vacuum space with $\s P_\ast(0)$ where $0$ is the zero-dimensional vector space.

These projectors are contained in the closure of $\r{Span}\bigl( \{\s E(X) \}\bigr)$ and as such inherit all relevant properties of the $\s E(X)$ listed in~(\ref{Eprops}).  They arise as the limits of normalized $\s E$-operators,
\begin{equation*}
\s P_\ast(\g K)
= \lim_{X_n \rightarrow [\g K]} \det(\idty - X_n)\ \s E \Bigl( \frac{X_n}{\idty - X_n} \Bigr), \qquad 0 \le X_n < \idty.
\end{equation*}

\begin{remark}
The map $\tilde{\s E}$ defined by
\begin{equation*}
\tilde{\s E} (X) \mapsto \det(\idty - X)\ \s E\Bigl( \frac{X}{\idty - X} \Bigr), \qquad \forall \, 0 \le X < \idty
\end{equation*}
is uniformly continuous and extends therefore continuously to $[0,\idty]$. This extension sheds some light on how the $n$-particle projectors arise as limits of $\s E$-operators.
\end{remark}

Suppose that $1$ is a $k$-degenerate eigenvalue of $X$. $X$ can then be decomposed as a direct sum of a projector P and a $(d-k)$-dimensional object:
\begin{equation*}
X = P \oplus \tilde{X}
\end{equation*}
For any sequence $0 \le (\epsilon_n)_n < 1$ that converges to one, the operators
\begin{equation*}
\tilde{\s E} ( \epsilon_n P \oplus \tilde X)
\end{equation*}
are well-defined, bounded and as per~(\ref{Eprops}) isomorphic to
\begin{equation*}
(1 - \epsilon_n)^k\, \det(\idty - \tilde X) \ \s E\Bigl(\frac{\epsilon_n}{1 - \epsilon_n}  P \Bigr) \otimes \s E\Bigl(\frac{\tilde X}{\idty - \tilde X}\Bigr).
\end{equation*}
By rearranging the factors in this expression, we get
\begin{equation*}
\tilde{\s E} ( \epsilon_n P) \otimes \tilde{\s E} (\tilde{X}).
\end{equation*}
We can write out the first factor as
\begin{equation*}
(1 - \epsilon_n)^k \Bigl\{ \Bigl( \loplus_{j\le k} (1-\epsilon_n)^{(k-j)} \epsilon_n^j \wedge^j P \Bigr) \loplus \Bigl( \loplus_{j>k} 0 \Bigr) \Bigr\},
\end{equation*}
and then it is clear that this will converge to
\begin{equation*}
0 \oplus \ldots 0 \oplus \bigl(\wedge^k P\bigr) \oplus 0 \ldots \oplus 0.
\end{equation*}

We will often ignore the possibility that $1$ is included in the spectrum of $X$ when we are talking about expressions containing $\tilde{\s E}(X)$.  When $1$ is contained in the spectrum of $X$, $f\bigl(\tilde{\s E} (X) \bigr)$ should be interpreted as
\begin{equation*}
\lim_{X_n \rightarrow X} f\bigl(\tilde{\s E} (X_n) \bigr), \qquad 0 \leq X_n < \idty
\end{equation*}
whenever $f$ is a continuous function.

\begin{remark} \label{rem:eigendecompEx}
We can express a general $\s E(X)$ using the above projectors as
\begin{equation*}
\s E(X) =  \sum_\Lambda x_\Lambda\, \s P_\ast\bigl( \g H_\Lambda \bigr)
\end{equation*}
where $\Lambda$ plays the same role as in~(\ref{Eprops}) and the $x_{\Lambda}$ are the products of the corresponding eigenvalues. So in essence, this gives us the eigendecomposition of $\s E(X)$.
\end{remark}

\section{Quasi-free States}
\label{sec:states}

The chief objects under study in this article are the so-called quasi-free states and maps. These states are the fermionic counterparts of gaussian measures
for classical systems or gaussian states for bosonic systems. They are
sometimes referred to as determinantal states or processes for reasons
that will soon become obvious.

A linear functional $\omega$ on the CAR
algebra which assigns zero values to all monomials in creation and
annihilation operators except for
\begin{equation*}
 \omega\bigl( a^*(\varphi_1) \cdots a^*(\varphi_k)
 a(\psi_k) \cdots a(\psi_1) \bigr)
 = \det\Bigl( \bigl[ \bigl\< \psi_i\, ,\, Q\,\varphi_j \bigr\> \bigr]
 \Bigr)
\end{equation*}
extends to a state on $\g A(\g H)$ if and only if the linear
one-particle space transformation $Q$ satisfies $0 \le Q \le \idty$.
Such an $\omega$ is called gauge invariant quasi-free and $Q$ its
corresponding \emph{symbol}. The notation $\omega_Q$ will be used to connect
the state to its symbol.

Using the language developed in \autoref{sec:GICAR} we can calculate the density matrix $\rho_Q$ corresponding to a state $\omega_Q$.

\begin{lemma}
The density matrix $\rho_Q$ corresponding to a state $\omega_Q$ with symbol $Q$ can be written down explicitly as
\begin{equation*}
\tilde{\s E} (Q) = \det(\idty - Q) \left\{ \idty \oplus \frac{Q}{\idty -Q} \oplus\left( \frac{Q}{\idty -Q} \wedge \frac{Q}{\idty -Q} \right)\oplus \ldots\right\}
\end{equation*}
\end{lemma}

\begin{proof}
Consider the symbol $Q$ of a general quasi-free state $\omega$. We can always find a 1D decomposition of $\g H$ such that it is amenable with the eigendecomposition of $Q$, i.e.\
\begin{equation*}
Q = \sum_i q_i \ket{e_i} \bra{e_i}, \qquad \g H_i = \Cx\, \ket{e_i}\, \& \, \oplus_i \g H_i  = \g H.
\end{equation*}
By straightforward computation we can check that $\omega_Q = \wedge_i \omega_{Q_i}$ with $Q_i := q_i |e_i\>\<e_i|$ and since
\begin{equation*}
\Gamma(\g H_1\oplus \g H_2\oplus \cdots\oplus \g H_n) = \Gamma(\g H_1) \otimes \Gamma(\g H_2) \otimes \cdots\otimes \Gamma(\g H_n),
\end{equation*}
$\g A(\oplus_i \g H_i) \cong \otimes_i \g A(\g H_i)$.

For the one-dimensional Hilbert space $\g H_i$, the Fock space $\Gamma(\g H_i)$ is 2-dimensional and the associated representation can be expressed on $\mathds{C}^2$ as
\begin{equation*}
\pi_{\Gamma} (a) = \begin{pmatrix} 0 &0 \\ 1 &0 \end{pmatrix}; \qquad
\ket{\Omega_F} = \begin{pmatrix} 0 \\ 1 \end{pmatrix}.
\end{equation*}
The relevant part of $Q$ on this Fock space is then
\begin{equation*}
Q_i = \begin{pmatrix} q &0 \\ 0 &1-q \end{pmatrix}.
\end{equation*}

Because of Proposition~\ref{prop:wedgestate} and the uniqueness implied in there, the wedge state $\wedge_i\, \omega_{Q_i}$ must be isomorphic to the product state
\begin{equation*}
\rho_{\tilde{Q}} = \rho_{Q_1} \otimes \ldots \otimes \rho_{Q_n}.
\end{equation*}
We can rewrite this in terms of exponential elements by
\begin{align*}
\rho_{\tilde Q}
&= \begin{pmatrix} q_1 &0 \\ 0 &1-q_1 \end{pmatrix} \otimes\cdots\otimes \begin{pmatrix} q_d &0 \\ 0 &1-q_d \end{pmatrix} \\
&= \det(\idty -Q)\, \begin{pmatrix} \frac{q_1}{1-q_1} &0 \\ 0 &1-q_1 \end{pmatrix} \otimes\cdots\otimes \begin{pmatrix} \frac{q_d}{1-q_d} &0 \\ 0 &1 \end{pmatrix},\qquad q_i \ne 1 \\
&= \det(\idty -Q)\, \s E\left(\frac{Q_1}{\idty-Q_1}\right) \otimes\cdots\otimes \s E\left(\frac{Q_d}{\idty - Q_d}\right) \\
\end{align*}
which is isomorphic to
\begin{equation*}
\rho_Q = \det(\idty-Q)\ \s E\left(\frac{Q}{\idty-Q}\right).
\end{equation*}
Finally, by continuity of the map $\tilde{\s E}$ this results also holds for $q_i=1$ and so in particular for projectors.
\end{proof}

By \hyperref[rem:eigendecompEx]{remark~\ref*{rem:eigendecompEx}}, there is an alternative way to write this density matrix.  Let $0 \le Q \le \idty$ be a linear transformation of the one-particle
space $\g H$ and let $\bigl\{ e_1, e_2, \ldots, e_d \bigr\}$ be an
orthonormal set of eigenvectors of $Q$, i.e.\
\begin{equation*}
 Q\, e_j = q_j\, e_j, \qquad 0 \le q_j \le 1,\ j=1,2,\ldots,d.
\end{equation*}
For a subset $\Lambda$ of $\{1,2,\ldots,d\}$ define
\begin{equation*}
 q_\Lambda := \prod_{r\in\Lambda} q_r\
 \prod_{s\in\{1,\ldots,d\}\setminus\Lambda} (1 - q_s)
 \qquad\text{and}\qquad
 \g H_\Lambda := \r{Span}\bigl( \bigl\{ e_r \,:\, r\in\Lambda \bigr\} \bigr)
\end{equation*}
then
\begin{equation*}
 \rho_Q = \sum_\Lambda q_\Lambda\, \s P_\ast\bigl( \g H_\Lambda \bigr).
\end{equation*}
As the $\s P_\ast(\g H_\Lambda)$ project on mutually orthogonal
subspaces, this decomposition shows that the $q_\Lambda$ are the
eigenvalues of $\rho_Q$. This allows to explicitly compute quantities
as the Renyi and von~Neumann entropies of $\rho_Q$.

As every $\s P_\ast(\g K)$ defines a pure state on $\g A(\g H)$ that is gauge
invariant and quasi-free, we see that every gauge invariant quasi-free
state is a convex mixture of pure gauge invariant quasi-free states.
The projectors $P_\ast$ are generally a small subset of the one-dimensional
projectors acting on $\g H^{(k)}$. This can be seen by a simple
parameter count. In order to parameterize a generic $m$-dimensional
complex subspace of $\Cx^n$, we need $2m(n-m)$ real parameters.
As $\dim\bigl( \g H^{(k)} \bigr) = \binom{d}{k}$ with $d = \dim(\g H)$
we need $2\bigl( \binom{d}{k} - 1 \bigr)$ real parameters to specify a
generic one-dimensional subspace of $\g H^{(k)}$ that is to say a
$k$-particle pure state on $\g A^{\r{GICAR}}(\g H)$ while we need only
$2k(d - k) \le 2\bigl( \binom{d}{k} - 1 \bigr)$ real parameters to specify
a $k$-dimensional subspace of $\g H$ which corresponds to a pure
quasi-free state. Therefore, the convex hull of the gauge invariant
quasi-free states is strictly smaller than the state space of
$\c A^{\r{GICAR}}(\g H)$. The linear span of the pure quasi-free states
coincides however with all linear functionals on $\c A^{\r{GICAR}}(\g
H)$.

To illustrate that the quasi-free states do not form a convex set on their own, we provide the following proposition which will also be of use later on.

\begin{proposition}
Let $\omega_{Q_1}$ and $\omega_{Q_2}$ be quasi-free and let
$0< \lambda < 1$, then $\lambda\, \omega_{Q_1} + (1-\lambda)\,
\omega_{Q_2}$ is quasi-free iff $Q_1 - Q_2$ is of rank 0 or 1.
Moreover, if the rank condition holds,
\begin{equation*}
 \lambda\, \omega_{Q_1} + (1-\lambda)\, \omega_{Q_2}
 = \omega_{\lambda\, Q_1 + (1-\lambda)\, Q_2}.
\end{equation*}
\end{proposition}
For the proof of this we refer to \cite{Wolfe1975}.

\subsection{Entropy Related Quantities of Quasi-free States}

The knowledge we have about the eigenvalue decomposition of a quasi-free state $\omega_Q$ allows us to translate the general expressions for entropy related quantities of that state to expressions on the one-particle density matrix. Especially calculations involving $p$-Renyi entropies become very simple in this way.

\begin{proposition}The $p$-Renyi entropy of a quasi-free state $\omega_Q$ is
\begin{equation}
\s H_{p} (\omega_Q) = \frac{1}{1-p}\, \tr \log \bigl( (\idty - Q)^p + Q^p \bigr).
\label{renyi}
\end{equation}
\end{proposition}

\begin{proof}
\begin{align*}
\s H_{p} (\rho_Q) :&= \frac{1}{1-p} \log \tr \rho_Q^p \\
&= \frac{1}{1-p} \log \left\{ \det \left( \idty + \left(\frac{Q}{\idty - Q} \right)^p \right) \det(\idty-Q)^p \right\} \\
&= \frac{1}{1-p} \log \det \bigl( (\idty - Q)^p + Q^p \bigr) \\
&= \frac{1}{1-p}\, \tr \log \bigl( (\idty - Q)^p + Q^p \bigr).
\end{align*}
\end{proof}

\begin{proposition}
The von Neumann entropy of a quasi-free state $\omega_Q$ is
\begin{equation*}
\s S(\omega_Q) = - \tr \Bigl( Q \log Q+ (\idty- Q) \log (\idty- Q) \Bigr).
\end{equation*}
\end{proposition}

\begin{proof}
The von~Neumann entropy is the limit of the $p$-Renyi entropy when $p\downarrow1$. As the expression in~(\ref{renyi}) becomes indeterminate (0/0) we use de~l'Hopital's rule combined with Jacobi's formula for differentiating a determinant
\begin{equation*}
 \frac{d\ }{dx} \det(A) = \tr\Bigl( A^{-1} \frac{dA}{dx} \Bigr)\, \det(A).
\end{equation*}
This yields
\begin{align*}
\s S(\omega_Q)
&= - \lim_{p\downarrow1} \frac{d\ }{dp}\, \log \det\left( (\idty - Q)^p + Q^p \right) \\
&= - \lim_{p\downarrow1} \frac{d\ }{dp}\, \det \bigl( (\idty - Q)^p + Q^p \bigr) \\
&= - \tr \Bigl( Q \log Q+ (\idty- Q) \log (\idty- Q) \Bigr).
\end{align*}
\end{proof}

\begin{proposition}
Let $Q_1$ and $Q_2$ be two symbols such that $\ker Q_2 \subset \ker Q_1$ and
$\ker(\idty-Q_2) \subset \ker(\idty-Q_1)$, then the relative entropy of $\omega_{Q_2}$ with respect to $\omega_{Q_1}$ is given by
\begin{equation*}
\begin{split}
\s S\bigl( \omega_{Q_1}; \omega_{Q_2} \bigr) = \tr \Bigl\{ &Q_1 \bigl( \log Q_1 - \log Q_2 \bigr) \\ &+ (\idty -Q_1) \bigl( \log (\idty - Q_1) - \log (\idty - Q_2) \bigr)\Bigr\}.
\end{split}
\end{equation*}
\end{proposition}

\begin{proof}
As above, the computation uses an appropriate limit
\begin{equation*}
\s S(\rho;\sigma) := \tr \rho(\log\rho - \log\sigma) = \lim_{p\to0} \frac{d\ }{dp}\, \tr\Bigl( \rho^{p+1} - \rho\,\sigma^p \Bigr).
\end{equation*}
Using~(\ref{renyi})
\begin{align*}
&\s S\bigl( \omega_{Q_1}; \omega_{Q_2} \bigr) \\
&\quad= \lim_{p\to0} \frac{d\ }{dp}\, \Bigl\{ \det(\idty - Q_1)^{p+1}\, \tr \s E\Bigl( \frac{Q_1^{p+1}}{(\idty - Q_1)^{p+1}} \Bigr) \\
&\quad\phantom{= \lim_{p\to0} \frac{d\ }{dp}\, \Bigl\{\ }- \det(\idty - Q_1)\, \det(\idty - Q_2)^p\, \tr \s E\Bigl( \frac{Q_1}{\idty - Q_1}\, \frac{Q_2^p}{(\idty - Q_2)^p} \Bigr)\Bigr\} \\
&\quad= \lim_{p\to0} \frac{d\ }{dp}\, \Bigl\{ \det\bigl( Q_1^{p+1}(\idty - Q_1)^{p+1}\bigr) - \det\bigl((\idty - Q_1)\, (\idty - Q_2)^p + Q_1 Q_2^p \bigr) \Bigr\} \\
&\quad= \tr \Bigl\{ Q_1 \bigl( \log Q_1 - \log Q_2 \bigr) + (\idty -Q_1) \bigl( \log (\idty - Q_1) - \log (\idty - Q_2) \bigr)\Bigr\}.
\end{align*}
\end{proof}

\section{Quasi-free Completely Positive Maps}
\label{sec:maps}

Let $\Lambda$ be a completely positive map on $\c M_d$. Dual to $\Lambda$ is another completely positive map $\Lambda^*$ on $\c M_d$
\begin{equation*}
\tr \Lambda(\sigma)\, X = \tr \sigma\, \Lambda^*(X),\quad \sigma,X \in \c M_d.
\end{equation*}
A general quantum operation may be described either in Schr\"odinger or in Heisenberg picture. In the first case we use completely positive maps $\Lambda$ with the additional property that $\tr \Lambda(\sigma) = \tr \sigma$. Such maps restrict to affine transformations of the state space of $\c M_d$ and are called trace-preserving completely positive (TPCP). Their duals $\Lambda^*$ leave the identity untouched and are therefore called unity-preserving completely positive (UPCP).

\subsection{Unital Maps}
\label{ssec:unital maps}

We will consider here two families of UPCP maps on $\g A(\g H)$ which generalize the expressions for quasi-free states \cite{Fannes1980}. Any unitary $U$ on one-particle space defines an automorphism of $\g A((\g H)$ through
\begin{equation*}
 a^*(\varphi) \mapsto a^*(U\,\varphi).
\end{equation*}
One can either check that the CAR are preserved or explicitly compute for an $m$-particle vector $\psi$
\begin{align*}
\s E(U) a^*(\varphi)\s E(U)^*\, \psi
&= \s E(U)a^*(\varphi)\, \bigl( \underbrace{U^* \otimes \cdots \otimes U^*}_{m \text{ times }}\, \psi\bigr) \\
&= \s E(U) \varphi \wedge \bigl( U^* \otimes \cdots \otimes U^*\, \psi\bigr) \\
&= \s E(U) \bigl(U^* \otimes\cdots\otimes U^* \bigr) \bigl( (U\, \varphi) \wedge \psi \bigr) \\
&= (U\, \varphi) \wedge \psi = a^*(U\, \varphi) \psi.
\end{align*}
Such automorphisms are called quasi-free.

Next consider a couple $A,B$ of linear transformations of $\g H$ such that
\begin{equation}
0 \le B \le \idty - A^*A.
\label{conqf}
\end{equation}
The block matrix
\begin{equation*}
V := \begin{pmatrix} A &\sqrt{\idty - AA^*} \\-\sqrt{\idty - A^*A} &A^* \end{pmatrix}
\end{equation*}
is unitary on $\g H \oplus \g H$ and, using the constraint on $(A,B)$, we can always find a linear map $0 \le Q \le \idty$ on $\g H$ such that $B = \sqrt{\idty - A^*A} Q \sqrt{\idty - A^*A}$. Using these ingredients, we construct a UPCP map $\Lambda^*_{A,B}$ on $\g A(\g H)$ by concatenating three UPCP maps: the injection $a^*(\varphi) \mapsto a^*(\varphi \oplus 0)$, the quasi-free automorphism defined by the block unitary $V$ and the projection $\id \wedge \omega_Q$ from $\g A(\g H \oplus \g H)$ to $\g A(\g H)$. An explicit computation yields
\begin{align*}
 &\Lambda^*_{A,B}\bigl( a^*(\varphi_1) \cdots a^*(\varphi_k)
 a(\psi_\ell) \cdots a(\psi_1) \bigr) \\
 &\quad=
 \sum \epsilon\, \omega_B\Bigl( a^*( \varphi_{m_1}) \cdots
 a^*( \varphi_{m_r}) a( \psi_{n_r}) \cdots
 a( \psi_{n_1}) \Bigr) \\
 &\phantom{\quad = \sum \epsilon} \times a^*( A\,\varphi_{i_1}) \cdots
 a^*(A\, \varphi_{i_{k-r}}) a( A\, \psi_{j_{\ell-r}}) \cdots
 a( A\, \psi_{j_1}).
\end{align*}
Here the summation is taken over all ordered partitions
\begin{equation*}
 \bigl\{ \{i_1,\ldots,i_{k-r}\}, \{m_1,\ldots,m_r\} \bigr\}
 \quad\text{and}\quad
 \bigl\{ \{j_1,\ldots,j_{\ell-r}\}, \{n_1,\ldots,n_r\} \bigr\}
\end{equation*}
of $\{1,\ldots,k\}$ and $\epsilon$ is the parity of the corresponding
permutation. It is not hard to show that the condition~(\ref{conqf}) is also necessary to have $\Lambda^*_{A,B}$ UPCP. This follows already from the requirement that any quasi-free state on $\g A(\g H)$ should be mapped into a state, which incidentally will also be quasi-free. From the definition one sees that the map $\Lambda^*$ restricts to a UPCP map of $\c A^{\r{GICAR}}$.

To define the second family, we first need a complex conjugation: fix an orthonormal basis $\{e_1, e_2, \ldots, e_d\}$ in $\Cx^d$. The elements of the basis will be considered as real vectors and the complex conjugation is the conjugate linear operator
\begin{equation*}
\varphi = \sum_j c_j\, e_j \mapsto \overline\varphi := \sum_j \overline{c_j}\, e_j .
\end{equation*}
From this definition we see that $\overline{(\overline\varphi)} = \varphi$ and $\<\overline\varphi \,,\, \overline\psi\> = \<\psi \,,\, \varphi\>$. We also introduce the conjugate of a complex linear transformation $A$ by $\overline A \varphi := \overline{A\,\overline\varphi}$. The transformation $\overline A$ is complex linear and satisfies
\begin{equation*}
\overline{A + \alpha B} = \overline A + \overline\alpha \overline B,\quad \overline{(\overline A)} = A,\quad \overline{A^*} = (\overline A)^*,\quad \text{and}\quad \overline{AB} = \overline A\, \overline B.
\end{equation*}
The entries of $\overline A$ in the distinguished basis $\{e_1, e_2, \ldots, e_d\}$ are
\begin{equation*}
(\overline A)_{ij} = \<e_i \,,\, \overline A e_j\> = \<e_i \,,\, \overline{A e_j}\> = \<A e_j \,,\, e_i\> = \overline{\<e_i \,,\, A e_j\>} = \overline{A_{ij}}.
\end{equation*}
In particular the conjugate coincides with the transpose for Hermitian elements.

The second family of maps we will consider is of the form
\begin{align*}
&\Gamma^*_{A,B}\bigl( a^*(\varphi_1) \cdots a^*(\varphi_k) a(\psi_\ell) \cdots a(\psi_1) \bigr) \\
&\quad=  \sum \epsilon\, \omega_B\Bigl( a^*( \varphi_{m_1}) \cdots  a^*( \varphi_{m_r}) a( \psi_{n_r}) \cdots  a( \psi_{n_1}) \Bigr) \\
&\phantom{\quad = \sum \epsilon} \times a( A\,\overline{\varphi_{i_1}}) \cdots  a(A\, \overline{\varphi_{i_{k-r}}}) a^*( A\, \overline{\psi_{j_{\ell-r}}}) \cdots  a^*( A\, \overline{\psi_{j_1}})
\end{align*}
with the same summation convention as above, $\Gamma_{A,B}^*$ is UPCP if and only if $0 \le B\le \idty - A^{\s T}(A^{\s T})^*$.

Both $\Lambda_{A,B}$ and $\Gamma_{A,B}$ map gauge-invariant quasi-free states into gauge-invariant quasi-free states
\begin{equation*}
\Lambda_{A,B}(\omega_Q) = \omega_{A^*QA+B} \quad\text{and}\quad \Gamma_{A,B}(\omega_Q) = \omega_{-A^{\s T}Q^{\s T}(A^{\s T})^* + B + A^{\s T} (A^{\s T})^*}.
\end{equation*}

We now compute the actions of $\Lambda^*_{A,B}$ and $\Gamma^*_{A,B}$ on elements of the type $\s E$.

\begin{lemma}
Suppose that $0 \le B \le \idty - A^*A$ and $\idty-B+XB$ invertible, then
\begin{align*}
\Lambda^*_{A,B}\bigl( \s E(X) \bigr)
&= \det(\idty - B + XB) \\
&\phantom{= \det(\idty - }\times \s E\Bigl( \idty + A (\idty -B+XB )^{-1} (X -\idty)A^*\Bigr) \\
&= \det(\idty - B + BX) \\
&\phantom{= \det(\idty -}\times \s E\Bigl( \idty + A (X -\idty)(\idty -B+BX )^{-1} A^*\Bigr).
\end{align*}
\end{lemma}

\begin{proof}
Suppose first that $0 \le B < \idty - A^*A$, the general case follows by continuity.  Choose now $0 \le Q < \idty$ and put $Q' := \gamma(Q) = A^*QA+B$, then also $0 \le Q' < \idty$. Recall also that the density matrix $\rho_Q = \det(\idty - Q)\, \s E(Q/(\idty-Q))$ (and the analogous expression for $\rho_{Q'}$). We now compute using
\begin{equation}
\det(\idty + CD) = \det(\idty + DC),
\label{det}
\end{equation}
\begin{align*}
& \tr \bigl( \Lambda_{A,B} (\rho_Q)\, \s E(X) \bigr) \\
&\quad= \det(\idty - A^*QA -B)\ \det\left( \idty + \frac{A^*QA +B}{\idty - A^*QA -B} X\right) \\
&\quad= \det\bigl( \idty - A^*QA -B + (A^*QA +B)X \bigr) \\
&\quad= \det\bigl( \idty -B + BX + A^*QA (X-\idty) \bigr) \\
&\quad= \det(\idty -B + BX)\ \det \bigl( \idty + A^* Q A (X-\idty)(\idty - B + BX)^{-1} \bigr)  \\
&\quad= \det(\idty -B + BX)\ \det \bigl( \idty + Q A (X-\idty)(\idty -B + BX)^{-1} A^* \bigr)  \\
&\quad= \det(\idty -B + BX)\ \det(\idty - Q)\\
&\phantom{\quad=\ }\times \det\Bigl( \frac{\idty}{\idty-Q} + \frac{Q}{\idty-Q} A (X-\idty) (\idty -B + BX)^{-1} A^* \Bigr)  \\
&\quad= \det\left(\idty +B(X - \idty)\right)\ \det(\idty - Q) \\
&\phantom{\quad=\ } \times \det\Bigl( \idty + \frac{Q}{\idty-Q} \bigl(\idty +  A (X-\idty) ( \idty -B + BX)^{-1} A^* \bigr) \Bigr) \\
&\quad= \tr \bigl( \rho_Q\, \Lambda^*_{A,B}(\s E(X)) \bigr).
\end{align*}
The second form of the expression follows from~(\ref{det}) and
\begin{equation*}
(\idty -B+XB )^{-1} (X -\idty) = (X -\idty)(\idty -B+BX )^{-1}.
\end{equation*}
\end{proof}

Although the previous theorem completely defines the map $\Lambda^*$, it might be useful to state how the action of this map looks on the density matrix of a quasi-free state.

\begin{proposition}
\label{prop6}
Let $Q$ be the symbol of a quasi-free state $\omega_Q$ and suppose that $\idty - Q + (2 Q - \idty)B$ is invertible, then
\begin{equation*}
\begin{split}
&\Lambda^*_{A,B}(\rho_Q) = \det\bigl(\idty - Q + (2 Q -\idty)B\bigr)
\nonumber\\
&\quad\times \s E\bigl( \idty + A (\idty - Q + (2 Q - \idty)B)^{-1} (2Q - \idty) A^* \bigr).
\end{split}
\end{equation*}
\end{proposition}

\begin{proof}
It suffices to replace $X$ with $\frac{Q}{\idty -Q}$ in the proof of the previous proposition and add the required normalization.  The result also remains valid when $Q$ becomes a projector, or even more generally, when $1 \in \sigma(Q)$ by the continuity of the map $\tilde{E}$.
\end{proof}

The maps $\Gamma^*$ can be handled in a similar way.

\begin{lemma}
Suppose that $0 \le B\le \idty - A^{\s T}(A^{\s T})^*$ and that $\idty -B^{\s T}- AA^* +X^{\s T}B^{\s T} + X^{\s T} AA^*$ is invertible, then
\begin{equation*}
\begin{split}
&\Gamma^*_{A,B}\bigl( \s E(X) \bigr) =
\det\bigl(\idty - B^{\s T} - AA^* + B^{\s T}X^{\s T} + AA^*X^{\s T} \bigr) \\
&\quad\times\ \s E\Bigl( \idty + A (\idty -B^{\s T}- AA^* +X^{\s T}B^{\s T} + X^{\s T} AA^* )^{-1} (\idty - X^{\s T}) A^*\Bigr).
\end{split}
\end{equation*}
\end{lemma}

\begin{proposition}
Let $Q$ be the symbol of a quasi-free state $\omega_Q$, $0 \le B\le \idty - A^{\s T}(A^{\s T})^*$ and suppose that $\idty - Q^{\s T} + (2 Q^{\s T} - \idty)(AA^*+B^{\s T})$ is invertible, then
\begin{equation*}
\begin{split}
&\Gamma^*_{A,B}(\rho_Q) = \det\bigl(\idty - Q^{\s T} + (2 Q^{\s T} -\idty)(AA^* + B^{\s T}) \bigr) \\
&\quad\times \s E\bigl( \idty + A \bigl(\idty - Q^{\s T} + (2 Q^{\s T} - \idty)(AA^*+B^{\s T}) \bigr)^{-1} (\idty - 2Q^{\s T}) A^* \bigr).
\end{split}
\end{equation*}
\end{proposition}

\subsection{Trace-preserving Maps}

The stability of the set of quasi-free states with respect to quasi-free CPTP maps can essentially be used as a characterization of such maps.

\begin{proposition}
The set of quasi-free states is invariant under a linear map $\Gamma$ if and only if the action of the map $\Gamma$ can be expressed as
\begin{equation} \label{qf-map}
\Gamma(\rho_Q) = \rho_{\gamma(Q)}
\end{equation}
where $\gamma(Q) = \pm A^* Q A + B$ or $\gamma(Q) = \pm A^* Q^{\s T} A + B$. Furthermore $\gamma(Q) = A^* Q A + B$ is CP iff $0 \le B \le \idty - A^*A$ and $\gamma(Q) = - A^* Q^{\s T} A + B$ is CP iff $A^*A \le B \le \idty$.
\end{proposition}

\begin{proof}
Consider a map\footnote{The $^*$ in this formula denotes the dual of the algebra, i.e. the span of the state space.} $\Gamma: (\g A^{GICAR})^* \rightarrow (\g A^{GICAR})^*$ with the following properties:
\begin{itemize}
\item trace-preserving
\item $\mathds{C}$-linear
\item maps quasi-free states onto quasi-free states
\end{itemize}

Since the map $\Gamma$ leaves the set of quasi-free states invariant, there must be a corresponding map on the set of symbols, so we propose a map $\gamma$ as in (\ref{qf-map}).  Consider now a matrix $Q$ which is an element of the open interval $[0,\idty[$ in $\g M(\mathds{C}^d)$ and a one-dimensional projector $P$ on the same algebra.  If $\epsilon$ is small enough (but non-zero) , $Q + \epsilon P$ is still an element of the unit interval.  Since the difference of $Q$ and $Q + \epsilon P$ is of Rank $1$
\begin{equation*}
\omega_{Q + \lambda \epsilon P} = \omega_{(1-\lambda) Q + \lambda (Q + \epsilon P)} = (1-\lambda) \omega_Q + \lambda \omega_{Q + \epsilon P}.
\end{equation*}
And because of the linearity of $\Gamma$, this gives us the following results about $\gamma$
\begin{align*}
& \r{Rank} \, \{ \gamma(Q + \epsilon P) - \gamma(Q) \} = 0 \quad \text{or} \quad 1, \qquad \forall \epsilon \\
\intertext{and}
&(1-\lambda) \gamma(Q) + \lambda \gamma(Q+ \epsilon P) = \gamma(Q + \lambda \epsilon P).
\end{align*}
We can rewrite this in a more useful form
\begin{align*}
\gamma(Q + \epsilon P) &= \frac{1}{\lambda} \{ \gamma(Q + \lambda \epsilon P) - (1-\lambda) \gamma(Q) \} \\
&= \epsilon \frac{1}{\lambda \epsilon} \{\gamma(Q + \lambda \epsilon P) - \gamma(Q) \} + \gamma(Q)\\
&= \gamma(Q) + \epsilon d(Q,P)
\end{align*}
where in the final line we have introduced the function $d(Q_1,Q_2)$ with the property that if $Q_2$ is a one-dimensional projector, the Rank of $d$ is $0$ or $1$.

As the maps $\Gamma$, $\tilde{\s E}$ and $\tilde{\s E}^{-1}$ are all Fréchet differentiable on the unit interval, $\gamma$ should also be differentiable.  The function $d$ we introduced above, is actually the derivative of $\gamma$ and as such it is linear in the second argument. So clearly for any matrix $0 \leq Q=\sum_i q_i P_i \leq \idty$ such that the $P_i$ are one-dimensional projectors,
\begin{equation*}
\gamma(Q)= \gamma(\sum_i q_i P_i)
= \sum_i q_i \, d(0,P_i) + \gamma(0)
\end{equation*}
Because of the Rank conditions on $d(Q,P)$ condition, we get
\begin{equation*}
\gamma(Q) = B \pm A^* Q A \qquad\text{or}\qquad \gamma(Q) = B \pm A^* Q^{\s T} A.
\end{equation*}
The conditions $0 \le B \le \idty - A^*A$ for the case $\gamma(Q) = A^* Q A + B$ and $A^*A \le B \le \idty$ for the case $\gamma(Q) = - A^* Q^{\s T} A + B$ are necessary because states have to be mapped onto states, in particular the image of a symbol should remain a symbol. Conversely, if the conditions hold, then we may define the dual UPCP maps as in \autoref{ssec:unital maps}
\end{proof}

\subsection{Jamio\l kowski states and Choi matrices of quasi-free completely positive maps}

There are two ways of encoding a CP map, in fact of encoding any super-operator, going under the names of Jamio\l kowski state and Choi matrix. Let $\{e_i\}$ be the standard basis of $\Cx^d$ with associated matrix units $e_{ij} := |e_i\>\<e_j|$. The Jamio\l kowski state $\s J(\Gamma)$ of
$\Gamma$ is defined as
\begin{equation*}
 \s J(\Gamma)
 := (\id \otimes \Gamma) \Bigl\{ {\textstyle\frac{1}{d}}\,
 \Bigl| \sum_i e_i \otimes e_i \Bigr\>
 \Bigl\< \sum_j e_j \otimes e_j \Bigr\> \Bigr|
 = {\textstyle\frac{1}{d}}\, \sum_{ij} e_{ij} \otimes \Gamma(e_{ij}).
\end{equation*}
It is easily verified that $\s J(\Gamma)$ is a state whenever $\Gamma$ is TPCP, i.e.\ a quantum operation in Schr\"odinger picture. The Choi matrix of a CP map $\Gamma$ is a very similar object
\begin{equation*}
 \s C(\Gamma)
 := \sum_{ij} e_{ij} \otimes \Gamma(e_{ij}),
\end{equation*}
mostly used in Heisenberg picture. So if $\Gamma$ is UPCP, the dual of a TPCP map, then the Choi matrix enjoys the property
\begin{equation*}
\tr_1 \s C(\Gamma) = \idty,
\end{equation*}
where $\tr_1$ denotes the partial trace over the first factor in $\Cx^d \otimes \Cx^d$.

We shall compute the Jamio\l kowski states of quasi-free TPCP maps $\Lambda_{A,B}$ and $\Gamma_{A,B}$ and the Choi matrices of their duals. Both the Jamio\l kowski state and the Choi matrix depend on the distinguished basis. A different choice of orthonormal basis in $\Cx^d$ yields however a unitarily equivalent state and matrix. We can therefore equally well define $\s J$ and $\s C$ with respect to a naturally chosen basis in fermionic context.

\subsubsection{Jamio\l kowski state of a TPCP map.}

\begin{proposition}
The Jamio\l kowski state of a quasi-free TPCP map $\Lambda_{A,B}$ or $\Gamma_{A,B}$ is unitarily equivalent to a gauge-invariant quasi-free with symbol
\begin{equation*}
\frac{1}{2} \begin{pmatrix} \idty &A \\ A^* &A^* A+2B \end{pmatrix}
\end{equation*}
for $\Lambda_{A,B}$ and
\begin{equation*}
\frac{1}{2} \begin{pmatrix} \idty &-A \\ -A^* &A^* A+2B^{\s T} \end{pmatrix}
\end{equation*}
for $\Gamma_{A,B}$.
\end{proposition}

\begin{proof}
Consider first the case of a TPCP map $\Lambda_{A,B}$. In order to stay within the context of gauge-invariant quasi-free states we embed $\g A(\g H)$ in the usual way in $\g A(\g H \oplus \g H)$, identifying $a(\varphi)$ with $a(0 \oplus \varphi)$. The gauge-invariant quasi-free state on $\g A(\g H \oplus \g H)$ with symbol
\begin{equation*}
{\textstyle\frac{1}{2}}\, \begin{pmatrix} \idty &\idty \\ \idty &\idty \end{pmatrix}
\end{equation*}
is pure and its marginal on $\g A\bigl( \{0\} \oplus \g H \bigr)$ is totally
mixed, hence it can be used to construct the Jamio\l kowski state.

The algebra $\g A(\g H \oplus \g H)$ can be decomposed as $\g A(\g H) \otimes \g A(\g H)$ but the factors cannot simply be chosen as $\g A(\g H \oplus 0)$ and $\g A(0 \oplus \g H)$, indeed, $a(\varphi_1 \oplus 0)$ anticommutes with $a(0 \oplus \varphi_2)$. There exists in $\g A(\g H)$ an element $\Theta$ such that $\Theta^* = \Theta$, $\Theta^2 = \idty$ and $\{\Theta \,,\, a(\varphi)\} = 0$. In fact, $\Theta$ is up to a sign uniquely defined by these requirements and using the Fock space representation of the CAR, we easily see that $\Theta = \pm \s E(-\idty)$. Let us now embed $\g A(\g H)$ in $\g A(\g H \oplus \g H)$ as $\imath(a(\varphi)) = \Theta_1\, a(0 \oplus \varphi)$ where $\Theta_1$ is the element in $\g A(\g H \oplus 0)$ just described. It is then easily checked that $\g A(\g H \oplus \g H)$ decomposes into the tensor product of $\g A(\g H \oplus 0)$ and the embedded algebra $\g A(\g H)$. Moreover,
\begin{align*}
\bigl( \id\otimes\Lambda^*_{A,B} \bigr) \bigl( a(\varphi_1 \oplus \varphi_2) \bigr)
&= \bigl( \id\otimes\Lambda^*_{A,B} \bigr) \bigl( a(\varphi_1 \oplus 0) + a(0 \oplus \varphi_2) \bigr) \\
&= \bigl( \id\otimes\Lambda^*_{A,B} \bigr) \bigl( a(\varphi_1 \oplus 0) + \Theta_1\, \Theta_1 a(0 \oplus \varphi_2) \bigr) \\
&= \bigl( \id\otimes\Lambda^*_{A,B} \bigr) \bigl( a(\varphi_1 \oplus 0) + \Theta_1 \otimes \imath\bigl( a(\varphi_2) \bigr) \bigr) \\
&= \bigl( a(\varphi_1 \oplus 0) + \Theta_1 \otimes \imath\bigl( a(A\,\varphi_2) \bigr) \bigr) \\
&= a(\varphi_1 \oplus A\,\varphi_2).
\end{align*}
In this way, we see that $\id \otimes \Lambda_{A,B}$ is again quasi-free on $\g A(\g H \oplus \g H)$ with defining operators
\begin{equation}
 \tilde A := \begin{pmatrix} \idty &0 \\ 0&A \end{pmatrix}
 \qquad\text{and}\qquad
 \tilde B := \begin{pmatrix} 0&0 \\ 0&B \end{pmatrix}.
\label{ext}
\end{equation}

It is now obvious that the Jamio\l kowski state will also be quasi-free with symbol
\begin{equation*}
\begin{split}
&\omega_{{\textstyle\frac{1}{2}}\,  \begin{pmatrix} \idty &\idty \\ \idty &\idty \end{pmatrix}} \Bigl\{ \bigl( \id \wedge \Lambda^*_{A,B} \bigr) \bigl( a^*(\varphi_1 \oplus \varphi_2) a(\psi_1 \oplus \psi_2)\bigr)\Bigr\} \\
&\quad= \Bigl\< \begin{pmatrix} \psi_1 \\ \psi_2 \end{pmatrix}\,,\, {\textstyle\frac{1}{2}}\, \begin{pmatrix} \idty &A \\ A^* &A^* A+2B \end{pmatrix}\, \begin{pmatrix} \varphi_1 \\ \varphi_2 \end{pmatrix}\Bigr\>.
\end{split}
\end{equation*}
Remark that the positivity conditions for the symbol of the Jamio\l kowski state precisely coincide with the positivity requirement $0 \le B \le \idty - A^*A$ for the TPCP map $\Lambda_{A,B}$.

The computation for a map $\Gamma_{A,B}$ is similar. It is now convenient to
compose $\id \wedge \Gamma_{A,B}$ with the local automorphism $\id \wedge \gamma$ where $\gamma(a(\varphi)) = a^*(\overline{\varphi})$. In this way we remain within the class of gauge-invariant quasi-free states on the composite algebra $\g A(\g H \oplus \g H)$. The extended map is now of the form $\Lambda_{\tilde A, \tilde B}$ with
\begin{equation*}
 \tilde A := \begin{pmatrix} \idty &0 \\ 0&(A^{\s T})^* \end{pmatrix}
 \qquad\text{and}\qquad
 \tilde B := \begin{pmatrix} 0&0 \\ 0&B \end{pmatrix}.
\end{equation*}
\end{proof}

\subsubsection{Choi matrix of a UPCP map.}

\begin{proposition}
The Choi matrix of a quasi-free UPCP map $\Lambda^*_{A,B}$ or $\Gamma^*_{A,B}$ is unitarily equivalent to
\begin{equation*}
\det(B)\ \s E\Bigl[ \begin{pmatrix} B^{-1} - \idty &B^{-1}A^* \\ AB^{-1} &\idty + AB^{-1}A^* \end{pmatrix} \Bigr]
\end{equation*}
for $\Lambda_{A,B}$ and
\begin{equation*}
\det(B)\ \s E\Bigl[ \begin{pmatrix} B^{-1} -\idty &B^{-1}A^{\s T} \\ (A^{\s T})^*B^{-1} &\idty + (A^{\s T})^*B^{-1}A^{\s T} \end{pmatrix} \Bigr]
\end{equation*}
for $\Gamma_{A,B}$.
\end{proposition}

\begin{proof}
On $\Cx^d \otimes \Cx^d$ the matrix $\sum_{ij} e_{ij} \otimes e_{ij}$ is equal to the dimension of the space times the projector on a maximally entangled vector. We should therefore compute
\begin{equation*}
2^d\, (\id \wedge \Lambda^*_{A,B}) \rho_Q,
\end{equation*}
where
\begin{equation*}
Q = {\textstyle\frac{1}{2}}\,  \begin{pmatrix} \idty &\idty \\ \idty &\idty \end{pmatrix}.
\end{equation*}
Using \hyperref[prop6]{Proposition~\ref*{prop6}} we obtain
\begin{align*}
&2^d \, \det(\idty - Q + (2Q - \idty)\tilde{B}) = \det\Bigl[ {\textstyle\frac{1}{2}}\, \begin{pmatrix} \idty &-\idty \\ 2B - \idty &\idty \end{pmatrix} \Bigr] = \det(B) \\
\intertext{and}
&\idty + \tilde A (\idty - Q + (2 Q - \idty)\tilde B)^{-1} (2Q - \idty) (\tilde A)^* = \begin{pmatrix} B^{-1} - \idty&B^{-1}A^* \\ AB^{-1} &\idty + AB^{-1}A^* \end{pmatrix},
\end{align*}
where $\tilde A$ and $\tilde B$ are as in~(\ref{ext}).

The computation for maps $\Gamma^*_{A,B}$ follows similar lines. Remark again that the positivity conditions for the Choi matrices coincide precisely with the conditions ensuring that $\Lambda_{A,B}$ and $\Gamma_{A,B}$ are CP.
\end{proof}

\bibliography{main}
\bibliographystyle{plain}

\end{document}